\documentstyle[prl,aps,epsfig]{revtex}

\begin{document}
\title{Competition between anisotropy and superconductivity in organic and cuprate
superconductors}
\author{T. Schneider}
\address{Physik-Institut der Universit\"{a}t Z\"{u}rich, Winterthurerstrasse 190,\\
CH-8057 Z\"{u}rich, Switzerland}
\maketitle

\begin{abstract}
We analyze the empirical correlation between the zero temperature
penetration depth $\lambda _{c}\left( 0\right) $ and the corresponding
normal state DC conductivity $\sigma _{c}^{DC}$, measured slightly above the
transition temperature $T_{c} $, in different classes of quasi
two-dimensional superconductors, including cuprates and organics. For this
purpose we invoke the scaling theory of quantum and finite temperature
critical phenomena. Important implications are: Superconductivity in the
organic and cuprate superconductors is a genuine three dimensional (3D)
phenomenon. The competition between anisotropy and superconductivity
destroys the latter in the 2D limit even in the ground state. The data
uncovers the flow to quantum criticality, including the 2D quantum
superconductor to insulator (2D-QSI) and the 3D quantum superconductor to
normal state (3D-QSN) transition. This flow gives a clear perspective of the
regimes where quantum fluctuations are essential and mean-field treatments
fail. Thus, a detailed account of the flow from mean-field to 2D-QSI
criticality in organics and the crossover from the 2D-QSI to the 3D-QSN
critical point in cuprates is a challenge for microscopic theories
attempting to solve the puzzle of superconductivity in these materials.
\end{abstract}

\bigskip

The formation of the superconducting condensate in elemental
metals and their alloys is well understood within the
BCS-Eliashberg mean-field theory of superconductivity. Here
pairing and phase coherence (superfluidity) set in at the same
temperature. Applicability to highly anisotropic
quasi-two-dimensional superconductors, including organic and
cuprate superconductors is challenged by the evidence for strong
thermal and quantum fluctuations in the normal and superconducting
state of these layered
materials\cite{book,fluc,shibauchime,clayton}. In this paper we
concentrate on interplane properties, emerging from the empirical
correlation between the zero temperature penetration depth
$\lambda _{c}\left( 0\right) $, due to supercurrents flowing
perpendicular to the layers, and the corresponding DC conductivity
$\sigma _{c}^{DC}\left( T_{c}^{+}\right) $, measured slightly
above the transition temperature $T_{c}$\cite{maple,dordevic}. To
provide an understanding of this empirical correlation and to
explore the implications, we invoke the scaling theory of finite
temperature and quantum critical phenomena. An essential role
plays the anisotropy. It is shown that the experimental data
points, in the $\left( \lambda _{c}\left( 0\right) ,\sigma
_{c}^{DC}\left( T_{c}^{+}\right) \right) $-plane, including
organic, cuprate and conventional superconductors, map the flow to
or away from limiting critical behavior. We identify the two
dimensional quantum superconductor to insulator (2D-QSI), the
three dimensional quantum superconductor to normal state (3D-QSN),
the finite temperature 3D-XY transition and its mean-field
counterpart as the essential limiting critical points. Since the
flow to 2D-QSI criticality implies increasing anisotropy, our
analysis reveals that superconductivity in the organics and
cuprates is a genuine 3D phenomenon and that the competition
between anisotropy and superconductivity destroys the latter in
the 2D limit even in the ground state. Moreover the flow to 2D-QSI
and 3D-QSN criticality and the associated crossovers give a clear
perspective of the regimes where quantum fluctuations are
essential and mean-field treatments fail. Thus, a detailed account
of the flow from mean-field to 2D-QSI criticality in organics and
the crossover from the 2D-QSI to the 3D-QSN critical point is a
challenge for microscopic theories attempting to solve the puzzle
of superconductivity in these materials. For a detailed discussion
of the generic phase diagram of cuprate superconductors and the
emerging critical points we refer to recent
reviews\cite{book,tschnei}.

\begin{figure}[tbp]
\centering
\includegraphics[totalheight=12cm]{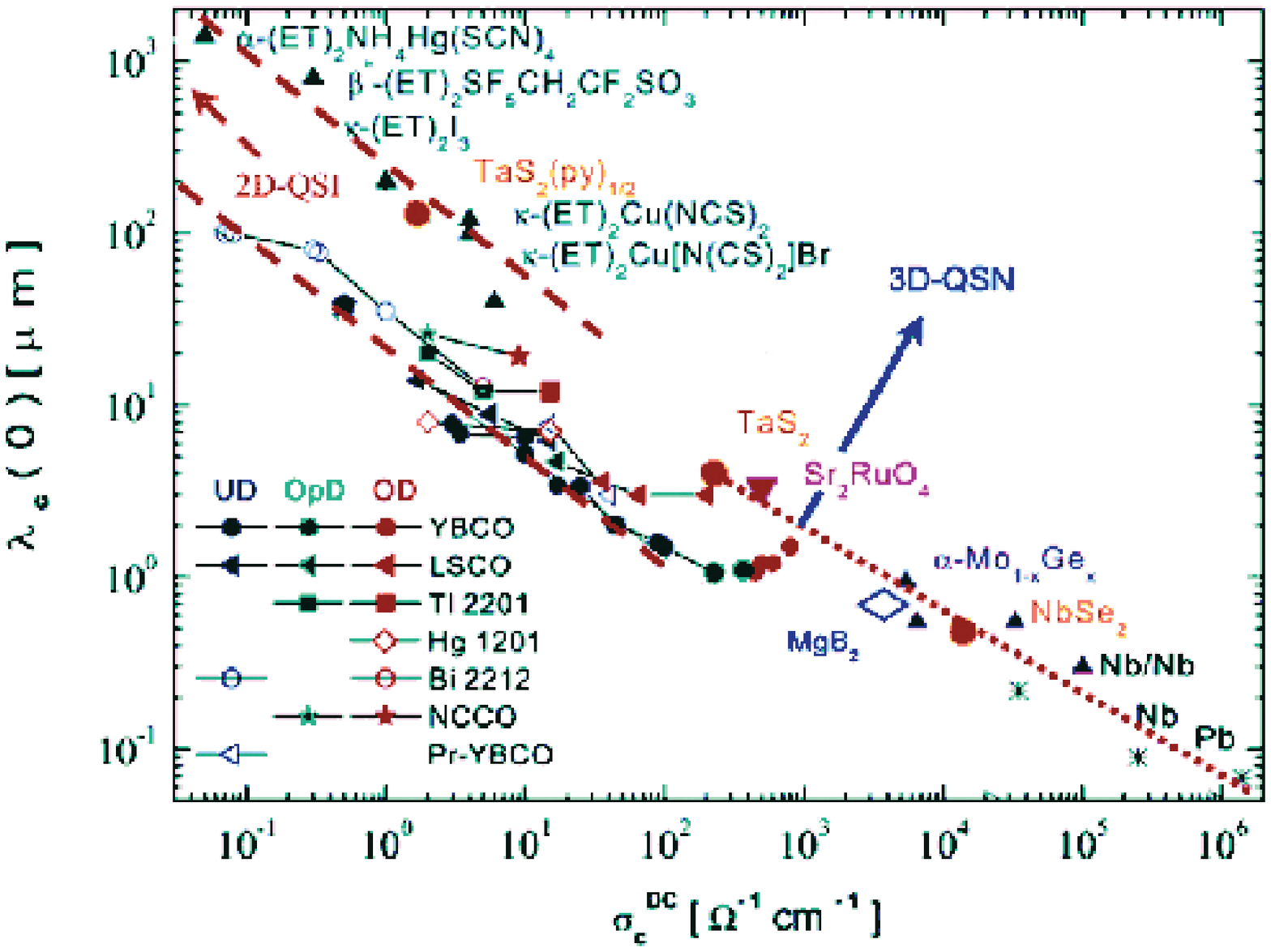}
\caption{$\protect\lambda _{c}\left( 0\right) $ versus $\protect\sigma %
_{c}^{DC}$ for a variety of superconductors. YBa$_{2}$Cu$_{3}$O$_{7-\protect%
\delta }$(YBCO) \protect\cite{ybco1,ybco2,ybco3,ybco4,ybco5}, La$_{2-x}$Sr$%
_{x}$CuO$_{4+\protect\delta }$ (LSCO)\protect\cite{ybco3,lsco6,lsco7}, HgBa$%
_{2}$CuO$_{4+\protect\delta }$(Hg1201)\protect\cite{hg}, Tl$_{2}$Ba$_{2}$CuO$%
_{6+\protect\delta }$ (Tl2201)\protect\cite{tl9,tl10,tl11}, Bi$_{2}$Sr$_{2}$%
CaCu$_{2}$O$_{8+\protect\delta }$ (Bi2212)\protect\cite{bi13,bi12}, Nd$%
_{2-x} $Ce$_{x}$CuO$_{4+\protect\delta }$ (NCCO)\protect\cite{nd14}. Blue
points -underdoped (UD), green -optimally doped (OpD) and red -overdoped
(OD). Transition metal dichalcogenides\protect\cite
{garoche,finley,trey,onabe,kennedy,thompson}; (ET)$_{2}$ X compounds
\protect\cite
{su1,carrington,shibauchi,dressel,taniguchi1,wanka,kajita,su2,prozorov}; Sr$%
_{2}$RuO$_{4}$\protect\cite{maeno,yoshida}; MgB$_{2}$\protect\cite
{mgb,angst,finnemore}; Nb \protect\cite{nb19,nb20}; Pb\protect\cite{nb20};
Nb Josephson junctions \protect\cite{nbjos}; $\protect\alpha -$Mo$_{1-x}$ Ge$%
_{x}$\protect\cite{moge}. The dotted line is Eq.(\ref{eq1}) and the dashed
ones Eq.(\ref{eq3}). The red arrows indicate the flow to 2D-QSI and for
overdoped YBa$_{2}$Cu$_{3}$O$_{7-\protect\delta }$ to 3D-QSN criticality,
respectively.}
\label{fig1}
\end{figure}

In Fig.\ref{fig1} we displayed $\lambda _{c}\left( 0\right) $ versus $\sigma
_{c}^{DC}\left( T_{c}^{+}\right) $ for a variety of cuprate, organic and
conventional superconductors, collected by Dordevic {\em et al. }\cite
{maple,dordevic} and Motohashi {\em et al.}\cite{bi13}. With the exception
of overdoped cuprates, we observe that the magnitude of $\lambda _{c}\left(
0\right) $ is systematically suppressed with the increase of the normal
state conductivity $\sigma _{c}^{DC}$. This feature appears to be consistent
with the relation
\begin{equation}
\lambda _{c}\left( 0\right) =\Omega _{s}\left( \sigma _{c}^{DC}\left(
T_{c}^{+}\right) \right) ^{-1/2},\ \ \Omega _{s}=\left( \frac{c^{2}\tau
_{tr}\left( T_{c}^{+}\right) }{4\pi }\right) ^{1/2},  \label{eq1}
\end{equation}
resulting from the standard expressions for the normal state conductivity $%
\sigma _{c}^{DC}\left( T\right) =ne^{2}\tau _{tr}\left( T\right) /m_{c}$ and
the definition of the London penetration depth $1/\lambda _{c}^{2}\left(
T=0\right) =\left( 4\pi ne^{2}\right) /\left( m_{c}c^{2}\right) $. $\tau
_{tr}$ is the mean scattering time of the electrons in transport properties
with number density n and effective mass $m_{c}$. This relationship also
agrees with the mean-field predictions for bulk superconductors in the dirty
limit\cite{tinkham} and layered BCS superconductors, treated as weakly
coupled Josephson junction\cite{bulaevski,ambegaokar,deutscher}, where $%
\Omega _{s}=\left( \hbar c^{2}/\left( 4\pi ^{2}\Delta \left( 0\right)
\right) \right) ^{1/2}$ and $\Delta \left( 0\right) $ denotes the zero
temperature energy gap. Thus, $\lambda _{c}\left( 0\right) \propto \left(
\sigma _{c}^{DC}\left( T_{c}^{+}\right) \right) ^{-1/2}$ is a characteristic
feature of conventional superconductors, where fluctuations do not play an
essential role. Indeed, the dotted line in Fig.\ref{fig1}, which is the
mean-field scaling form (\ref{eq1}), captures the trend emerging from the
data of conventional superconductors, as well as of MgB$_{2}$ and Sr$_{2}$RuO%
$_{4}$, remarkably well. Prominent and systematic deviations occur for the
highly anisotropic organic\ and underdoped cuprate superconductors. As $%
\lambda _{c}\left( 0\right) $ increases with the reduction of the normal
state conductivity $\sigma _{c}^{DC}$, we observe that the data tends to
fall onto two branches, consistent with
\begin{equation}
\lambda _{c}\left( 0\right) =\Omega _{s}\left( \sigma _{c}^{DC}\right)
^{-x},x\approx 3/4.  \label{eq3}
\end{equation}
The most prominent deviations from these trends occur for overdoped
cuprates. In particular the data for overdoped YBa$_{2}$Cu$_{3}$O$_{7-\delta
}$(YBCO) exhibit an upturn, suggesting a crossover to a power law of the
form (\ref{eq3}) but with a negative exponent $x$. Note that this crossover
appears to be correlated with the disappearance of the pseudogap in the
overdoped regime.

An essential ingredient for an understanding of the empirical correlations
between $\lambda _{c}\left( 0\right) $ and $\sigma _{c}^{DC}$ is the
anisotropy. For cuprates it is well established that doping and substitution
affect the anisotropy $\gamma $ dramatically\cite{book,tschnei}. For
cuprates it is well established that doping and substitution affect the
anisotropy $\gamma $ dramatically\cite{book,tschnei}. In tetragonal cuprates
it is defined as the ratio $\gamma =\xi _{ab}/\xi _{c}$ of the correlation
lengths parallel ($\xi _{ab}$) and perpendicular ($\xi _{c}$) to CuO$_{2}$
layers (ab-planes). In the superconducting state it can also be expressed as
the ratio $\gamma =\lambda _{c}/\lambda _{ab}$ of the London penetration
depths due to supercurrents flowing perpendicular ($\lambda _{c}$) and
parallel ($\lambda _{ab}$) to the ab-planes. Approaching a nonsuperconductor
to superconductor transition $\xi $ diverges, while in a superconductor to
nonsuperconductor transition $\lambda $ tends to infinity. In both cases,
however, $\gamma $ remains finite as long as the system exhibits anisotropic
but genuine 3D behavior. There are two limiting cases: $\gamma =1$
characterizes isotropic 3D- and $\gamma =\infty $ 2D-critical behavior. An
instructive model where $\gamma $ can be varied continuously is the
anisotropic 2D Ising model \cite{onsager}. When the coupling in the y
direction goes to zero, $\gamma =\xi _{x}/\xi _{y}$ becomes infinite, the
model reduces to the 1D case and $T_{c}$ vanishes. In the Ginzburg-Landau
description of layered superconductors the anisotropy is related to the
interlayer coupling. The weaker this coupling is, the larger $\gamma $ is.
The limit $\gamma =\infty $ is attained when the bulk superconductor
corresponds to a stack of independent slabs of thickness $d_{s}$. With
respect to experimental work, a considerable amount of data is available on
the chemical composition dependence of $\gamma $. At $T_{c}$ it can be
inferred from resistivity ($\gamma =\xi _{ab}/\xi _{c}=\sqrt{\rho _{c}/\rho
_{ab}}$) and magnetic torque measurements, while in the superconducting
state it follows from magnetic torque and penetration depth ($\gamma
=\lambda _{c}/\lambda _{ab}$) data. Experimentally it is known that as the
dopant concentration is reduced, $\gamma _{T_{c}}$ and $\gamma _{T=0}$
increase systematically, and tend to diverge in the underdoped limit. Here $%
T_{c}$ vanishes and the cuprates undergo a 2D-QSI transition, dominated by
quantum fluctuations. For this reason the scaling theory of critical
phenomena should apply. At finite temperatures and in the normal state it
predicts\cite{book,tschnei}
\begin{equation}
\sigma _{ab}^{DC}\propto \frac{\xi _{\tau }}{\xi _{c}},\ \sigma
_{c}^{DC}\propto \frac{\xi _{c}\xi _{\tau }}{\xi _{ab}^{2}},  \label{eq4}
\end{equation}
where $\xi _{\tau }$ is the correlation length associated with the critical
dynamics. At $T_{c}$ the ratio is then simply given by the anisotropy
\begin{equation}
\frac{\sigma _{c}^{DC}}{\sigma _{ab}^{DC}}=\left( \frac{\xi _{c}}{\xi _{ab}}%
\right) ^{2}=\left( \frac{1}{\gamma _{T_{c}}}\right) ^{2}.  \label{eq5}
\end{equation}
On the other hand, at zero temperature the penetration depths and anisotropy
are related by
\begin{equation}
\lambda _{c}\left( 0\right) =\gamma _{T=0}\lambda _{ab}\left( 0\right) .
\label{eq6}
\end{equation}
Close to 2D-QSI criticality, the scaling theory of quantum critical
phenomena applies\cite{book,tschnei,kim}. Here $T_{c}$, $\gamma _{T}$ and $%
\lambda _{ab}\left( 0\right) $ scale as\cite{book,tschnei,tskelis}
\begin{equation}
T_{c}\propto \left( \lambda _{ab}\left( 0\right) \right) ^{-2}\propto \delta
^{\overline{\nu }z},\ \ \gamma _{T_{c}}=\gamma _{T_{c},0}\delta ^{-\overline{%
\nu }},\ \gamma _{T=0}=\gamma _{0,0}\delta ^{-\overline{\nu }},\lambda
_{ab}\left( 0\right) =\lambda _{ab,0}\left( 0\right) \delta ^{-\overline{%
z\nu }/2}  \label{eq7}
\end{equation}
where $\overline{\nu }$ is the correlation length, $z$ the dynamic critical
exponent of the 2D-QSI transition and $\delta $ measures the distance from
this critical point. $\gamma _{T_{c},0}$, $\gamma _{0,0}$ and $\lambda
_{ab,0}\left( 0\right) $ denote the critical amplitudes close to 2D-QSI
criticality. Another essential property of this critical point is that for
any finite $T_{c}$ the in-plane areal conductivity is always larger than\cite
{book,kim}
\begin{equation}
\sigma _{ab}^{DC}d_{s}\approx \frac{4e^{2}}{h},  \label{eq8}
\end{equation}
where $d_{s}$ is the thickness of the independent sheets. Thus close to
2D-QSI criticality the normal state D.C. conductivity $\sigma _{c}^{DC}$,
evaluated above but close to $T_{c}$, and zero temperature penetration depth
$\lambda _{c}\left( 0\right) $ scale then as
\begin{equation}
\lambda _{c}\left( 0\right) =\Omega _{s}\ \sigma _{c}\left( T_{c}^{+}\right)
^{-\left( 2+z\right) /4},\ \ \Omega _{s}\approx \gamma _{0,0}\lambda
_{ab,0}\left( 0\right) \left( \frac{4e^{2}}{\gamma _{T_{c},0}d_{s}h}\right)
^{\left( 2+z\right) /4}  \label{eq9}
\end{equation}
In cuprates there is mounting evidence for a 2D-QSI transition with $z=1$%
\cite{book,tskelis,klosters}. This value is also consistent with the
theoretical prediction for a disordered bosonic system with long-range
Coulomb interaction. Here the loss of superfluidity is due to the
localization of the pairs , which is ultimately responsible for the
transition\cite{fisher,herbut}. The resulting exponent $\left( 2+z\right)
/2=3/4$ coincides with that in Eq.(\ref{eq3}), yielding the dashed lines in
Fig.\ref{fig1}. Since organic superconductors undergo superconductor to
insulator transitions\cite{nakazawa}, are highly anisotropic and $\gamma $
can be varied over a rather extended interval, i.e. $\gamma _{T_{c}}=180$
for $\kappa $-(ET)$_{2}$ Cu[N(CS)$_{2}$ ]Br, $\gamma _{T_{c}}=350$ for $%
\kappa $-(ET)$_{2}$ Cu(NCS)$_{2}$\cite{kawamata} and $\gamma _{T_{c}}=2\
10^{3}$ for $\alpha $- (BEDT-TTF)$_{2}$NH$_{4}$Hg(SCN)$_{4}$\cite{taniguchi2}%
, it becomes clear that the organics line in Fig.\ref{fig1} marks the flow
to 2D-QSI criticality, where the scaling relation (\ref{eq9}) applies, as
well. This interpretation is also confirmed by the empirical linear
relationship between $T_{c}$ and $1/\lambda _{ab}^{2}\left( 0\right) $ for
various cuprate and organic superconductors\cite{uemura,uemurabed},
consistent with the scaling relation (\ref{eq7}). It is a characteristic
property of nearly two dimensional systems\cite{kim,book}. An instructive
example is the onset of superfluidity in $^{4}$He films adsorbed on
disordered substrates, where the linear relationship between $T_{c}$ and
areal superfluid density is well confirmed \cite{crowell}. Moreover, the
existence of a threshold in-plane conductivity $\sigma _{ab}^{DC}d_{s}$ (Eq.(%
\ref{eq8})) is also well documented for Zn-substituted La$_{2-x}$Sr$_{x}$CuO$%
_{4}$ and YBa$_{2}$Cu$_{3}$O$_{7-\delta }$\cite{fukuzumi} and underdoped YBa$%
_{2}$Cu$_{3}$O$_{7-\delta }$\cite{semba}. What remains to be shown is wether
or not Eq.(\ref{eq9}) yields for $\Omega _{s}$ the correct order of
magnitude. Considering La$_{2-x}$Sr$_{x}$CuO$_{4}$, where $\gamma
_{0,0}\approx 1.63$, $\gamma _{T_{c},0}\approx 2$, $d_{s}\approx 6.6\ A$ and
$\lambda _{ab,0}\left( 0\right) \approx 740\ A$\cite{tschnei} we obtain for $%
z=1$, $\Omega _{s}\approx 24\mu m\left( \Omega cm\right) ^{3/4}$, which is
reasonably close to $\Omega _{s}\approx 18\mu m\left( \Omega cm\right) ^{3/4}
$, the value of the cuprates line in Fig.\ref{fig1}, which matches the data
points of underdoped La$_{2-x}$Sr$_{x}$CuO$_{4}$. Unlike layered BCS
superconductors where $\Omega _{s}$ is determined by the gap and therefore
by $T_{c}$ (Eq.(\ref{eq2})), this scale is mainly fixed by the critical
amplitude $\lambda _{ab,0}\left( 0\right) $ of the zero temperature in-plane
penetration depth and the effective thickness of the layers $d_{s}$ at
2D-QSI criticality. For this reason $\Omega _{s}$ does not adopt a universal
value. Thus taken together, the interpretation that the dashed lines in Fig.%
\ref{fig1} mark the flow of underdoped cuprate and organic superconductors
to 2D-QSI criticality is well established. From the relationship
\begin{equation}
T_{c}\propto \left( \frac{1}{\lambda _{ab}\left( 0\right) }\right)
^{2}\propto \left( \frac{1}{\lambda _{c}\left( 0\right) }\right) ^{2z/\left(
2+z\right) }\propto \left( \sigma _{c}^{DC}\left( T_{c}^{+}\right) \right)
^{z/2}\propto \gamma _{T_{c}}^{-z}\propto \gamma _{T=0}^{-z},  \label{eq10}
\end{equation}
derived from Eqs.(\ref{eq5}) to (\ref{eq9}), it then follows that $T_{c}$
and $\sigma _{c}^{DC}\left( T_{c}^{+}\right) $ vanish, while $\lambda
_{c}\left( 0\right) $, $\lambda _{ab}\left( 0\right) $ and the anisotropy $%
\gamma $ diverge at 2D-QSI criticality. It is clear from this scaling
relation how superconductivity competes with anisotropy and the associated
enhancement of fluctuations. It implies that in organic and underdoped
cuprate superconductors, a no vanishing transition temperature and
superfluid density in the ground state is unalterably linked to a finite
anisotropy. This raises serious doubts that 2D models are potential
candidates to explain superconductivity in these materials. Moreover, since $%
\sigma _{ab}^{DC}$ is always larger than $4e^{2}/hd_{s}$ (Eq.(\ref{eq8})),
the observed generic upturn of $\rho _{c}\left( T\right) $ in underdoped
cuprates\cite{fukuzumi,semba}, is according to Eq.(\ref{eq5}), yielding $%
\rho _{c}\left( T_{c}^{+}\right) =\gamma _{T_{c}}^{2}/\sigma
_{ab}^{DC}\left( T_{c}^{+}\right) \approx \gamma _{T_{c}}^{2}h/4e^{2}$, a
3D-2D crossover phenomenon and with that a characteristic signature of the
flow to 2D-QSI criticality.

Next we turn to the behavior of overdoped cuprates. Experimentally it is
known that in the overdoped limit $T_{c}$ tends to zero, the anisotropy $%
\gamma $ remains finite, while $\lambda _{ab}\left( 0\right) $ and with that
$\lambda _{c}\left( 0\right) $ diverges. This follows from the $\mu $SR data
for Y$_{0.8}$Ca$_{0.2}$Ba$_{2}$(Cu$_{1-y}$Zn$_{y}$)O$_{7-\delta }$ (Y$_{0.8}$%
Ca$_{0.2}$-123), Tl$_{0.5-y}$Pb$_{0.5+y}$Sr$_{2}$Ca$_{1-x}$Y$_{x}$Cu$_{2}$O$%
_{7}$(Tl-1212)\cite{bernhard} and TlBa$_{2}$CuO$_{6+\delta }\ $(Tl-2201)\cite
{niedermayer}, where the heavily overdoped regime was attained. Since $%
\gamma $ remains finite a 3D quantum superconductor to normal state (QSN)
transition occurs. $\xi _{\tau }$ scales as $\xi _{\tau }\propto \xi
_{ab}^{z_{cl}}$, where $z_{cl}$ is the critical exponent of the finite
temperature dynamics. In this case the scaling theory predicts\cite
{book,tschnei}
\begin{equation}
\sigma _{c}^{DC}\propto \frac{\xi _{c}\xi _{\tau }}{\xi _{ab}^{2}}\propto
\frac{\xi _{\tau }}{\xi _{ab}\gamma _{T_{c}}}\propto \xi
_{ab}^{z_{cl}-1}\propto \delta ^{-\overline{\nu }\left( z_{cl}-1\right) },\
\ \lambda _{c}\left( 0\right) \propto \delta ^{-\frac{\overline{\nu }}{2}%
\left( 1+z\right) },  \label{eq11}
\end{equation}
so that
\begin{equation}
\lambda _{c}\left( 0\right) \propto \left( \sigma _{c}\left(
T_{c}^{+}\right) \right) ^{\frac{1+z}{2\left( z_{cl}-1\right) }}\propto
T_{c}^{-\frac{z_{cl}-1}{z}}  \label{eq12}
\end{equation}
A potential candidate for the 3D-QSN transition is the Ginzburg-Landau
theory proposed by Herbut \cite{herbutd}. It describes the disordered d-wave
superconductor to disordered metal transition at weak coupling and is
characterized by the critical exponents $z=2$ and $\overline{\nu }=1/2$,
except in an exponentially narrow region. Although the overdoped cuprates
included in Fig.\ref{fig1} are still quite far from the overdoped limit, the
upturn of the data for slightly overdoped YBa$_{2}$Cu$_{3}$O$_{7-\delta }$%
(YBCO), signals with the scaling relation (\ref{eq12}) the flow to 3D-QSN
criticality with $z_{cl}>1$, as indicated in Fig.\ref{fig1}. Thus unlike the
organic superconductors the cuprates undergo a doping tuned crossover from
2D-QSI to 3D-QSN criticality, where $\sigma _{c}^{DC}\left( T_{c}^{+}\right)
$ and $\lambda _{c}\left( 0\right) $ tend to infinity, while $T_{c}$
vanishes. These critical points are attained in the underdoped and overdoped
limit, respectively, where $T_{c}$ vanishes. In this crossover cuprates pass
either to the organics line or the dotted one, spanned by conventional
superconductors (see Fig.\ref{fig1}. The YBa$_{2}$Cu$_{3}$O$_{7-\delta }$%
(YBCO) and La$_{2-x}$Sr$_{x}$CuO$_{4+\delta }$ (LSCO) \ cross the dotted
mean-field line in the neighborhood of TaS$_{2}$, Sr$_{2}$RuO$_{4}$ and MgB$%
_{2}$. Since Sr$_{2}$RuO$_{4}$ exhibits Fermi liquid normal state properties
\cite{bergemann}, fluctuations do not play an essential role in this region
of the $\lambda _{c}\left( 0\right) $-$\sigma _{c}^{DC}\left(
T_{c}^{+}\right) $ plane. In the heavily overdoped regime, however, the flow
to the 3D-QSN transition (red arrow)\ enhances quantum fluctuations and
invalidates mean-field concepts.\ On the contrary, the organic
superconductors evolve rather smoothly from the mean-field regime, to which
Pb, Nb, Nb-junctions, $\alpha $-Mo$_{1-x}$Ge$_{x}$ and the dichalcogenides
belong, to 2D-QSI criticality. Thus, the flow to 2D-QSI and 3D-QSN
criticality and the associated crossovers give a clear perspective of the
regimes where quantum fluctuations are essential. Since the order parameter
is a complex scalar, the finite temperature critical behavior falls into the
3D-XY universality class\cite{book,tschnei}. Due to the existence of the
2D-QSI and 3D-QSN critical points, and of the mean-field line, its detection
will be hampered by the associated crossovers which reduce the temperature
regime where thermal 3D-XY fluctuations dominate. Nevertheless, our analysis
clearly revealed that superconductivity in the organic and cuprate
superconductors is a genuine 3D phenomenon and that the competition between
anisotropy and superconductivity destroys the latter in 2D limit, even in \
the ground state. When the aforementioned universality classes of the 2D-QSI
and 3D-QSN transitions hold true, disorder plays an essential role. At
2D-QSI criticality it localizes the pairs and destroys superfluidity\cite
{fisher,herbut} and at the 3D-QSN transition it destroys, combined with a
weakening of the pairing interaction, superfluidity and pairs\cite{herbutd}.
A detailed account of the flow from mean-field to 2D-QSI criticality in
organics and the crossover from the 2D-QSI \ to the 3D-QSN critical point is
a challenge for microscopic theories attempting to solve the puzzle of
superconductivity in these materials. \bigskip

\acknowledgments The author is grateful to A.R. Bishop, H.Keller, K.A.
M\"{u}ller and J. Roos for suggestions on the subject matter. 

\end{document}